\documentclass[%
 reprint,
 superscriptaddress,
 amsmath,amssymb,
 aps,
]{revtex4-1}

\usepackage[utf8x]{inputenc}
\usepackage{graphicx}
\usepackage{dcolumn}
\usepackage{bm}
\usepackage{textgreek}

\usepackage{bm,mathrsfs,dcolumn,graphicx,color}
\definecolor{ablue}{rgb}{0.1,0.35,0.75}
\definecolor{agreen}{rgb}{0,0.6,0.3}
\definecolor{ared}{rgb}{0.4,0,0}

\usepackage[normalem]{ulem}

\begin{document}


\title{Exciton propagation and halo formation in two-dimensional materials}

\author{Raül Perea-Causín}
\author{Samuel Brem}
\author{Roberto Rosati}
\author{Roland Jago}
\affiliation{Department of Physics, Chalmers University of Technology, 412 96 Gothenburg, Sweden}
\author{Marvin Kulig}
\author{Jonas D. Ziegler}
\author{Jonas Zipfel}
\author{Alexey Chernikov}
\affiliation{Department of Physics, University of Regensburg, Regensburg D-93053, Germany}
\author{Ermin Malic}
\affiliation{Department of Physics, Chalmers University of Technology, 412 96 Gothenburg, Sweden}


\begin{abstract}
The interplay of optics, dynamics and transport is crucial for the design of novel optoelectronic devices, such as photodetectors and solar cells. In this context, transition metal dichalcogenides (TMDs) have received much attention. Here, strongly bound excitons dominate optical excitation, carrier dynamics and diffusion processes. While the first two have been intensively studied, there is a lack of fundamental understanding of non-equilibrium phenomena associated with exciton transport that is of central importance e.g. for high efficiency light harvesting. In this work, we provide microscopic insights into the interplay of exciton propagation and many-particle interactions in TMDs. Based on a fully quantum mechanical approach and in excellent agreement with photoluminescence measurements, we show that Auger recombination and emission of hot phonons act as a heating mechanism giving rise to strong spatial gradients in excitonic temperature. The resulting thermal drift leads to an unconventional exciton diffusion characterized by spatial exciton halos.
\end{abstract}

\maketitle


\noindent\textbf{\large{Introduction}}\\[5pt]
Exciton dynamics in atomically thin transition metal dichalcogenides (TMDs) has been intensively studied, tracking the way of excitons in time and momentum\,\cite{mueller2018exciton, wang2018colloquium, chernikov2015population, steinhoff2016nonequilibrium, selig2018dark, brem2018exciton, Merkl2019ultrafast}. In contrast,  their spatio-temporal dynamics including exciton propagation remained only little explored. In addition to the initial reports\,\cite{Kumar2014a,Mouri2014,Yuan2017,Kato2016}, more recent experimental studies on exciton diffusion in TMDs show efficient exciton transport at room temperature\,\cite{cadiz2018exciton} as well as strong density dependence of effective diffusion and the formation of intriguing spatial rings (halos) at elevated excitation densities\,\cite{kulig2018exciton}.
The latter phenomena indicate strong non-equilibrium effects in the spatial dynamics of optical excitations in two-dimensional materials, motivating the necessity to develop fundamental understanding of the underlying elementary processes.
Microscopic insights into the intriguing non-linear exciton diffusion and its interplay with many-particle scattering processes are thus of central importance to allow to understand, predict, and control exciton transport towards high-performance optoelectronics devices.

\begin{figure}[t!]
 \centering
 \includegraphics[width=\linewidth]{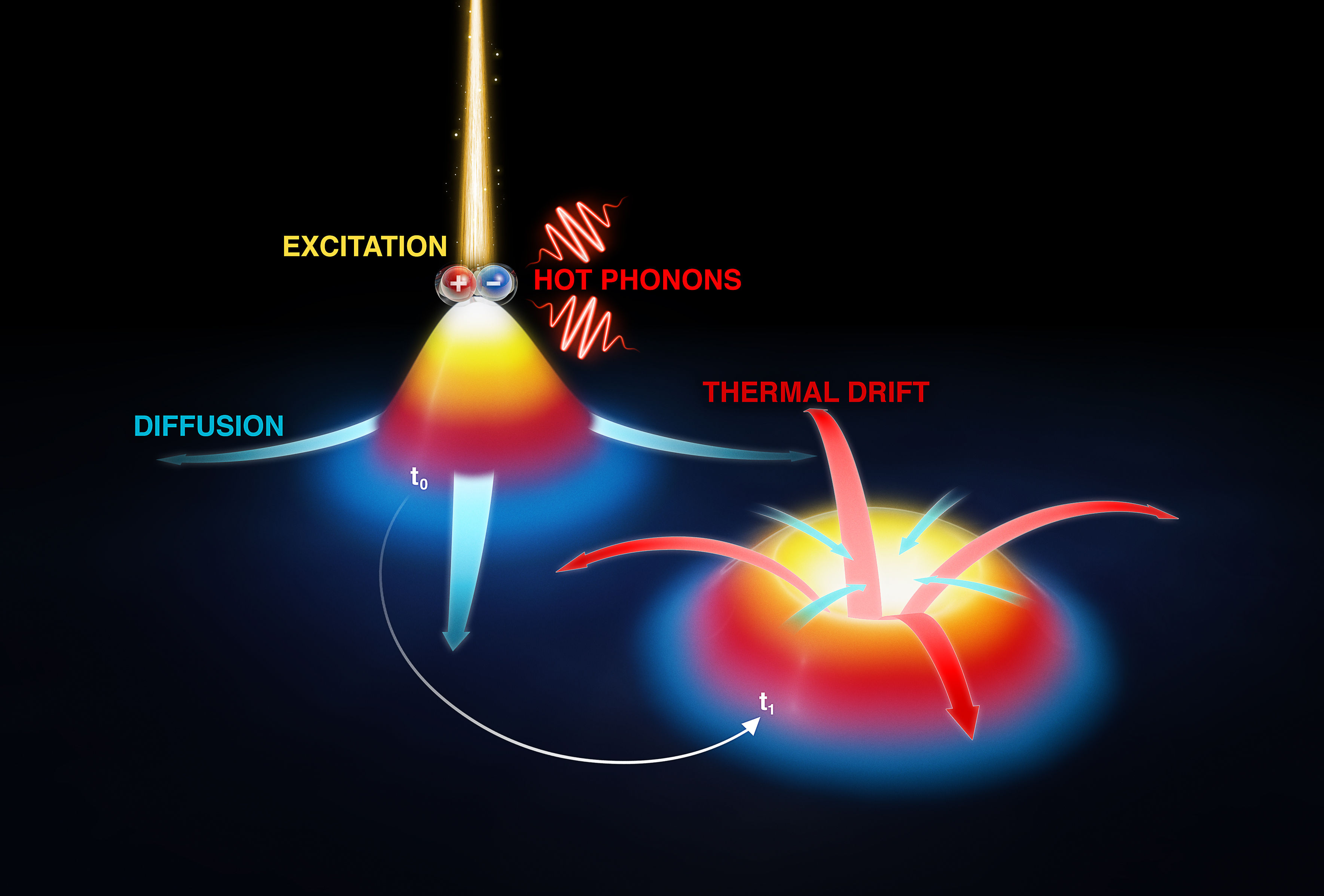}
 \caption{\textbf{Exciton propagation and halo formation}. Excitons are locally generated by an optical excitation and start diffusing to homogenize the spatial distribution ($t_0\rightarrow t_1$). Through Auger recombination and relaxation by hot-phonon emission, a long-lived spatial gradient in excitonic temperature is formed (color gradient). This results in a thermal drift dragging excitons out of the central region and giving rise to the formation of spatial rings (halos).}
 \label{fig:schema}
\end{figure}

In this work, our main goal is to provide a fundamental understanding of the interplay between optics, exciton dynamics and exciton transport in TMDs. We apply a microscopic approach based on the density matrix formalism\,\cite{rossi2002theory, kira2006many, malic2013graphene} to describe the spatio-temporal exciton and phonon dynamics. Our calculations  quantitatively reproduce temporally and spectrally resolved photoluminescence measurements of non-trivial exciton propagation at elevated excitation conditions.  We are thus able to study the fundamental mechanisms behind the exciton diffusion and investigate the impact of exciton-phonon scattering processes beyond the standard bath approximation. We reveal that in the strong excitation regime hot-phonon effects and, as a consequence, hot-exciton phenomena become crucial. The phonons heat up the exciton system in the central spatial region giving rise to a super-Gaussian exciton diffusion and to the formation of halo-like profiles in the spatial exciton distribution (cf. Fig.~\ref{fig:schema}). To explain this non-linear diffusion, we derive a generalized Fick's law to account for additional thermal drift currents (Seebeck effect) appearing as a direct consequence of the strong spatial gradients in the exciton temperature. A direct comparison to experimentally measured photoluminescence spectra shows an excellent agreement with respect to the evolution of halo-like spatial profiles as well as their formation time at different excitations.\\


\noindent\textbf{\large{Results}}\\[5pt]
Using the density matrix formalism, equations of motion for the spatio-temporal dynamics of excitons and phonons are derived by exploiting the Heisenberg equation and the many-particle Hamilton operator\,\cite{kira2006many, haug2009quantum, malic2013graphene}. The dynamics of phonons is explicitly considered to account for hot-phonon emission and reabsorption beyond the standard bath approximation. The derived equations are then transformed into the Wigner representation\,\cite{hess1996maxwell, jago2019spatio} (cf. methods section) in order to have access to the space-, time- and energy-resolved dynamics of excitons $N_{\mathbf{Q}}(\mathbf{r}, t)$ and phonons $n_{\mathbf{q}}(\mathbf{r}, t)$:
\begin{align}
\dot{N}_{\mathbf{Q}}(\mathbf{r}, t) &= - \mathbf{v}_{\mathbf{Q}} \cdot \mathbf{\nabla}_{\mathbf{r}} N_{\mathbf{Q}}(\mathbf{r}, t)- r_A N_{\mathbf{Q}}(\mathbf{r},t) N(\mathbf{r},t) \label{eq:EOM_f} \\
 &\quad + \Gamma^{ \text{in}}_{\mathbf{Q}}(\mathbf{r}, t) - \Gamma^{\text{out}}_{\mathbf{Q}}(\mathbf{r}, t) N_{\mathbf{Q}}(\mathbf{r}, t) \nonumber \\[0.2cm]
 \dot{n}_{\mathbf{q}}(\mathbf{r}, t) &= \Gamma^{\text{em}}_{\mathbf{q}}(\mathbf{r}, t) \left(1+n_{\mathbf{q}}(\mathbf{r}, t)\right) - \Gamma^{\text{abs}}_{\mathbf{q}}(\mathbf{r}, t) n_{\mathbf{q}}(\mathbf{r}, t) \nonumber \\ &\quad + \alpha N^2(\mathbf{r},t) - \kappa \left( n_{\mathbf{q}}(\mathbf{r}) - n^{\text{0}}_{\mathbf{q}} \right) \label{eq:EOM_n}
\end{align}
with $\mathbf{v}_{\mathbf{Q}} = \pm \hbar \mathbf{Q} M_X^{-1}$ as the exciton velocity, $\kappa$ as the phonon decay rate, $n^{0}_{\mathbf{q}}$ as the equilibrium phonon occupation (described by a Bose distribution), and the exciton-phonon scattering rates $\Gamma^{i}_{\mathbf{q}}(\mathbf{r}, t)$ being computed consistently according to the second-order Born-Markov approximation\,\cite{selig2018dark, brem2018exciton}. Here we consider scattering of excitons with the most efficient phonon modes including longitudinal and transverse acoustic and optical modes and the out-of-plane A$_1$ mode\,\cite{jin2014intrinsic}.

The first term and the entire second line in Eq.~\eqref{eq:EOM_f} account for exciton propagation and exciton-phonon scattering, respectively. While these interactions have been introduced on a fully  microscopic footing, the second term describing the Coulomb-induced Auger scattering is treated on a semi-phenomenological level. We describe the decrease of exciton population $N$ due to exciton-Auger processes with the experimentally accessible coefficient $r_A=0.4$ cm$^2/$s\,\cite{kulig2018exciton}, defined according to the density-dependent recombination rate $N r_A$. This term accounts for the decay of one exciton with the momentum $\mathbf{Q}$ and the excitation of another exciton to a higher energetic state. The latter then relaxes towards the ground state by emitting a cascade of optical phonons (LO, TO and A$_1$ modes at the \textGamma-point).
The corresponding optical phonon emission rate due to Auger recombination enters as a source term in the equation for phonons (first term in the second line in Eq.~\eqref{eq:EOM_n}). The appearing rate  $\alpha = 2 \pi \varepsilon_{\text{X}} r_A (\varepsilon_{\text{op}} q_c^2)^{-1}$ is determined by the Auger recombination rate $r_A$, the ratio between the energy of the exciton $\varepsilon_{\text{X}}$ and optical phonons $\varepsilon_{\text{op}}$, and the cut-off momentum $q_c$ for the emission of optical phonons.
\\


\begin{figure}[t!]
 \centering
 \includegraphics[width=\linewidth]{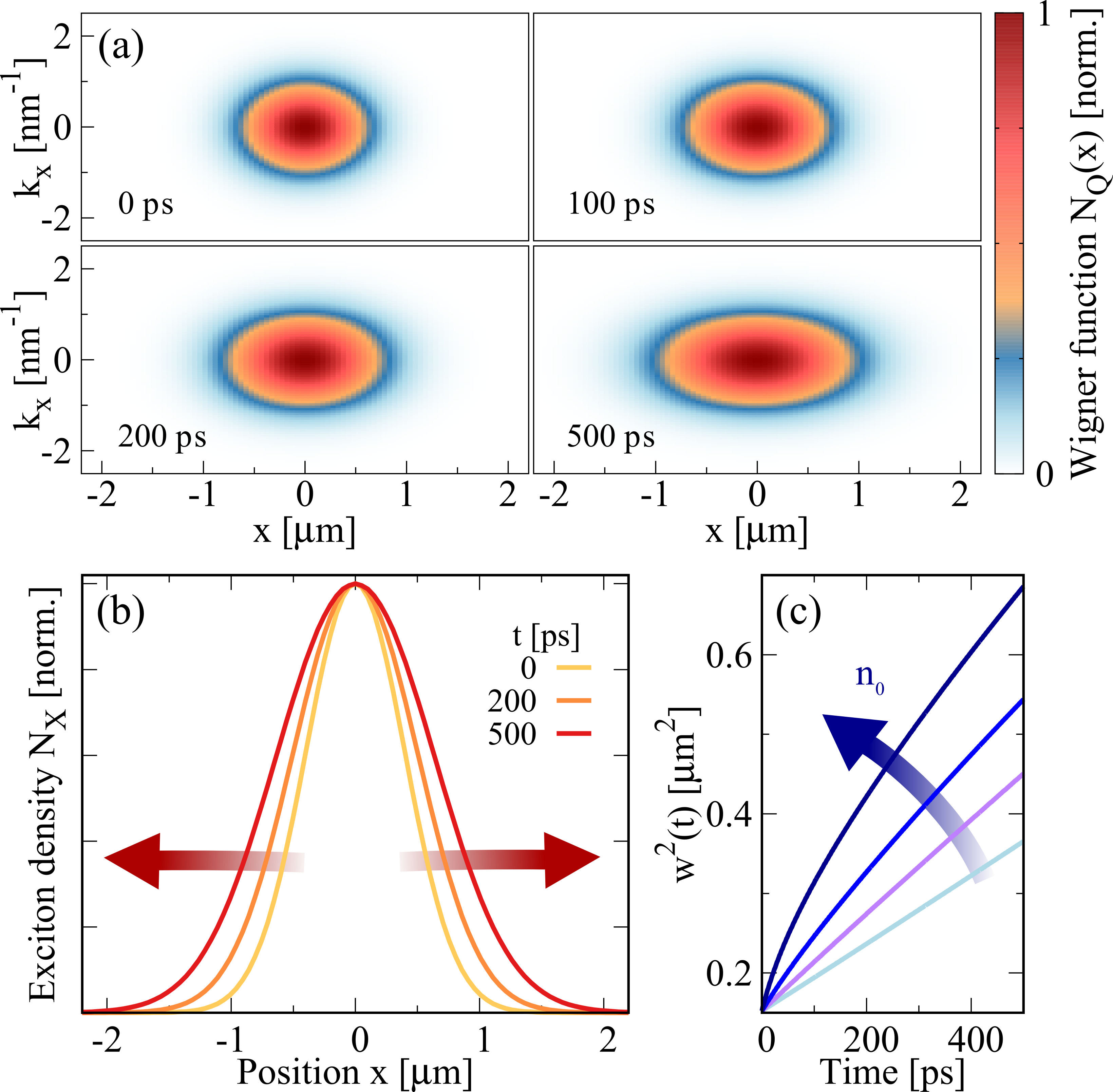}
 \caption{\textbf{Exciton diffusion in low-excitation regime}. (a) Phase-space representation of the normalized exciton Wigner function at different times for an initial peak exciton density of $10^9\,\text{cm}^{-2}$. (b) Spatial distribution of the exciton density at different times. (c) Spatial variance of the exciton density for different initial densities ($10^9$, $10^{10}$, $3 \cdot 10^{10}$, and $10^{11}\,\text{cm}^{-2}$). }
 \label{fig:LinearDiffusion}
\end{figure}

\noindent\textbf{Conventional exciton diffusion}\\
Solving the equations of motion, we have microscopic access to the diffusion of excitons in arbitrary TMD materials. Here, we consider the exemplary case of WS$_2$ monolayers on a SiO$_2$ substrate at room temperature. The material-specific input parameters for the electronic band structure (effective mass, band gap), carrier-phonon coupling elements, and phonon decay rate were taken from Ref.\,\onlinecite{kormanyos2015k, jin2014intrinsic, kulig2018exciton}. 

We start by solving  the coupled equations of motion for exciton and phonon occupations in the low-excitation regime, i.e. for exciton densities in the range of $10^9\,\text{cm}^{-2}$. 
These densities are sufficiently high to provide thermodynamically stable exciton population above the threshold for entropy ionization\,\cite{Mock1978,Steinhoff2017}, due to high binding energies on the order of $300\,\text{meV}$.
Figure~\ref{fig:LinearDiffusion} shows the exciton Wigner function in real- and momentum-space along the x-coordinate for four fixed times. Here, excitons quickly thermalize in reciprocal space through scattering with phonons and propagate in real space with their given momentum resulting in a broadening of the spatial distribution. This is further visualized in the spatial profiles of the exciton density (Fig.~\ref{fig:LinearDiffusion} (b)), illustrating a clear broadening of the exciton distribution. 

In the considered low-excitation regime, the exciton distribution keeps at all times its initial Gaussian shape -- in agreement with the solution of Fick's laws of diffusion for the exciton current 
$
 \mathbf{j}(\mathbf{r},t) = - D \mathbf{\nabla} N(\mathbf{r},t) $ with the diffusion coefficient $D$ and the exciton density $N(\mathbf{r}, t)$ \cite{PATHRIA2011583}. The latter is determined using the continuity equation
$ \dot{N}(\mathbf{r},t) = - \mathbf{\nabla} \cdot \mathbf{j}(\mathbf{r},t) = D \nabla^2 N(\mathbf{r},t) $ with the solution
$ N(\mathbf{r},t) = \frac{N_0}{4 \pi D t} \exp{\big(- \frac{\mathbf{r}^2}{w^2(t)}\big)}
$. Here, the spatial variance of the exciton distribution $w^2(t) =  \big< r^2(t) \big>= w_0^2 + 4 D t$ shows a clear linear time dependence  (Fig.~\ref{fig:LinearDiffusion} (c)), where the slope determines the exciton diffusion coefficient $D$ in the linear regime. At elevated densities, however, the extracted slope increases with the excitation density, ranging from values for an effective diffusion coefficient $D_{\text{eff}}$ of $2\,\text{cm}^2\,\text{s}^{-1}$ in the low-excitation regime up to $20\,\text{cm}^2\,\text{s}^{-1}$ at an enhanced density of $10^{12}\,\text{cm}^{-2}$. Note that for increasing excitation densities, the spatial variance of the exciton density is no longer linear in time (Fig.~\ref{fig:LinearDiffusion} (c)). The fact that its slope becomes smaller suggests that excitons move and recombine faster in the excited high-density region and then slow down and exhibit longer lifetimes once they leave the region. As further discussed below, a dominant aspect of this behavior is a long-lived spatial temperature gradient in the exciton distribution that is due to the emission and reabsorption of hot phonons.\\

\begin{figure}[t!]
 \centering
 \includegraphics[width=\linewidth]{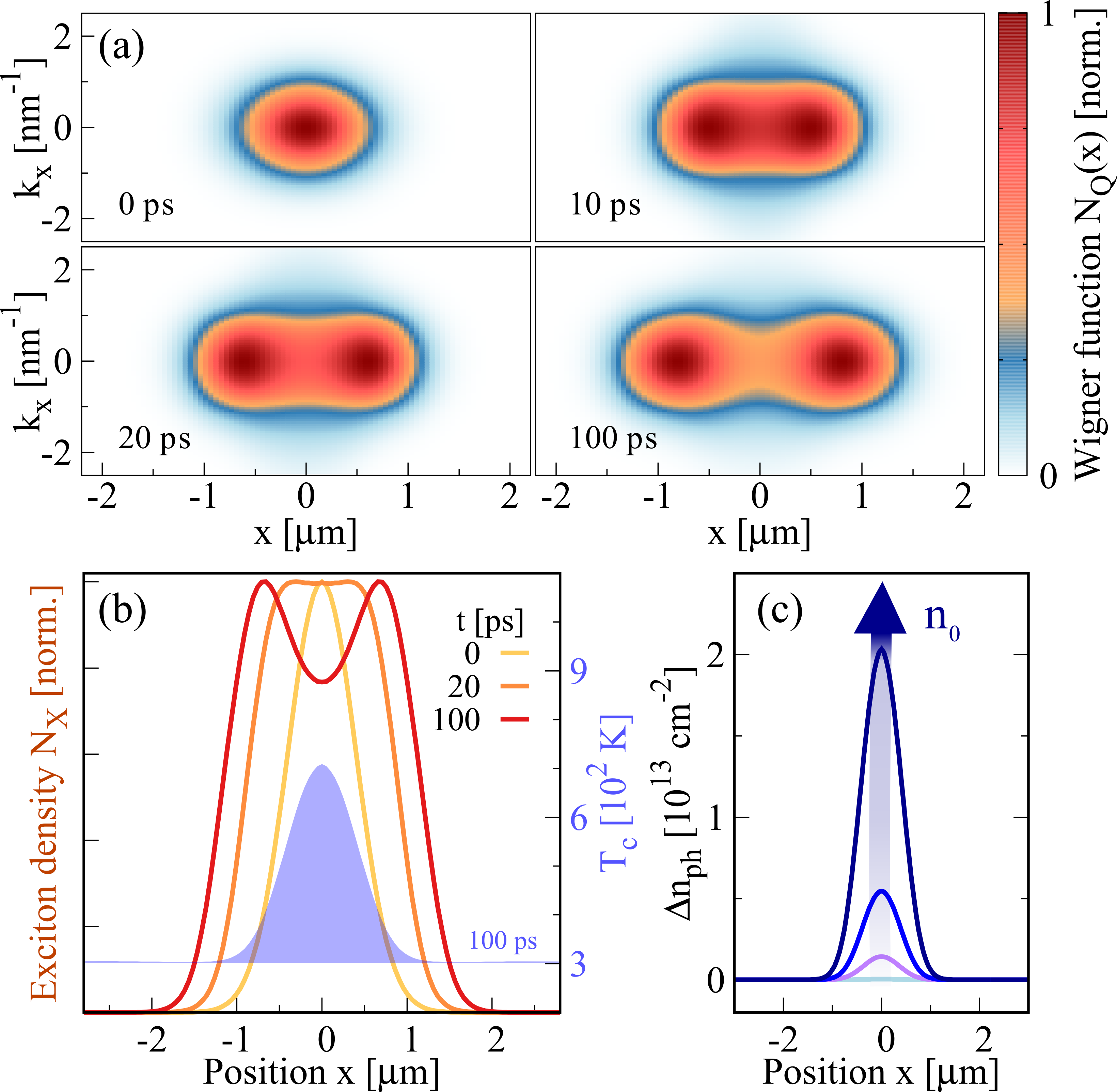}
 \caption{\textbf{Non-linear exciton diffusion and halo formation in the high-excitation regime}. (a) Phase-space representation of the normalized exciton Wigner function at different times for an initial peak exciton density of $n_0 = 10^{12}\,\text{cm}^{-2}$. (b) Spatial distribution of the exciton density at different times along the exciton temperature at 100 ps after the excitation (blue shaded curve). (c) Spatial profile of the excess optical phonon density at 100 ps for different initial densities ($10^{10}$, $10^{11}$, $3 \cdot 10^{11}$, and $10^{12}\,\text{cm}^{-2}$).}
 \label{fig:nonlinearDiffusion}
\end{figure}

\noindent\textbf{Non-linear exciton diffusion and halo formation}\\
To investigate the origin of the density-dependent boost in the diffusion coefficient and the non-linear behavior of the exciton spatial variance, we study now the diffusion in the high-excitation regime considering an initial exciton density of $10^{12}\,\text{cm}^{-2}$. This density is sufficiently high for the exciton-exciton interactions to become very efficient, yet remains still at least an order of magnitude below the ionization threshold for the Mott transition\,\cite{Chernikov2015c,Steinhoff2017}. We show again the space- and momentum-dependent Wigner function for four fixed times (Fig.~\ref{fig:nonlinearDiffusion} (a)). In contrast to the low-excitation case, we observe a few picoseconds after the excitation an apparent accumulation of excitons at low momenta outside the central region. However, the spatial dip in the center of the Wigner function is a result of a broader distribution in momentum space (corresponding to a higher effective temperature).
The resulting temperature gradient induces an efficient thermal drift that leads to the depletion of the central region, as will be discussed later.
For the spatially resolved exciton density (Fig.~\ref{fig:nonlinearDiffusion} (b)), this corresponds to the evolution of the initial Gaussian profile into a super-Gaussian distribution promptly followed by the appearance of spatial rings (halos) around the central region. Similar ring-shaped emission patterns have also been observed in GaAs-based quantum wells\,\cite{Butov2002}.  

The observed non-linear diffusion is induced by emission and reabsorption of hot optical phonons, i.e. the optical phonon distribution significantly deviates from the equilibrium Bose distribution. Note that although Auger recombination alone can contribute to the formation of super-Gaussian spatial profiles, it cannot explain the formation of halos. Figure~\ref{fig:nonlinearDiffusion} (c) illustrates the excess phonon density that  becomes very large at the central excited spatial region  - strongly depending on the excitation density $n_0$. A direct consequence is the occurrence of a pronounced spatial gradient in the effective excitonic temperature, cf. the blue shaded curve in Fig.~\ref{fig:nonlinearDiffusion} (b), which we have determined by using the equipartition theorem $ N^{-1}(\mathbf{r}, t) \sum_{\mathbf{Q}} \varepsilon_{\mathbf{Q}} N_{\mathbf{Q}}(\mathbf{r},t) = k_B T_X(\mathbf{r},t)$.
During the relaxation of Auger-scattered excitons, a large number of optical phonons is emitted, which are then reabsorbed heating up the excitons in the central excited region. The higher the excitation density, the more pronounced is this effect, leading eventually to a non-linear exciton diffusion that cannot be described anymore by the standard Fick's law of diffusion. 

This standard law can be generalized to account for thermal effects having a crucial impact on exciton diffusion. An expression for the spatially dependent current $\mathbf{j}$ due to gradients in the exciton density $N$ and the exciton temperature $T_X$ can be derived within a relaxation time approximation for the scattering, introducing the exciton relaxation time $\tau_{\mathbf{Q}} = (\Gamma^{\text{in}}_{\mathbf{Q}} + \Gamma^{\text{out}}_{\mathbf{Q}})^{-1}$. Assuming that diffusion creates a very small deviation $N_{\mathbf{Q}}^1$ from the equilibrium distribution $N_{\mathbf{Q}}^0$ and that this deviation is stationary on the timescale of diffusion and that its spatial gradients are weak, we can find an expression for $N_{\mathbf{Q}}^1$. By integrating the velocity of this excess distribution, we obtain the current due to spatial and thermal gradients:
\begin{equation}
 \mathbf{j}(\mathbf{r},t) = - D \mathbf{\nabla}_{\mathbf{r}} N(\mathbf{r},t) - \sigma s \mathbf{\nabla}_{\mathbf{r}} T_X(\mathbf{r},t) \label{eq:j_nT} 
 \end{equation}
 with  the conductivity $\sigma$ and the Seebeck coefficient $s$, where
 $\sigma s = (2 k_B T_X^2)^{-1} \sum_{\mathbf{Q}} \tau_{\mathbf{Q}} v_{\mathbf{Q}}^2 \left( \varepsilon_{\mathbf{Q}} - k_B T_X \right) N^0_{\mathbf{Q}} \approx \tau k_B N M_X^{-1}$. The appearing diffusion coefficient 
 reads
$ D = (2 N)^{-1} \sum_{\mathbf{Q}} \tau_{\mathbf{Q}} v_{\mathbf{Q}}^2 N^0_{\mathbf{Q}} \approx \tau k_B T_X M_X^{-1}
 $. Note that in contrast to the standard Fick's law, an additional second term appears in Eq.~\eqref{eq:j_nT}. This term corresponds to thermal drift (Seebeck effect) and accounts for excitons moving from hot towards colder regions.
 
Applying Eq.~\eqref{eq:j_nT} to the continuity equation yields a modified law of diffusion that takes into account spatial gradients in the exciton temperature. Here, excitons diffuse in space to flatten and eventually remove the initially introduced spatial non-uniformity in the exciton density and in the temperature profiles. When a strong temperature gradient is created due to the substantial hot-phonon emission, thermal drift becomes an important contribution to the current. A sufficiently strong temperature gradient is able to drag excitons out of the hot region faster than diffusion, leading first to super-Gaussian profiles and eventually to the appearance of spatial rings, as shown in Fig.~\ref{fig:nonlinearDiffusion}. Some excitons will then diffuse back from the rings into the hot region in order to counteract this additional, spatially non-uniform distribution of the population. The calculated and observed halo formation, however, demonstrates that the thermal drift is much more efficient than diffusion in the strong excitation regime for the studied system (cf. Fig.~\ref{fig:schema}).\\ 

\begin{figure}[t!]
 \centering
 \includegraphics[width=\linewidth]{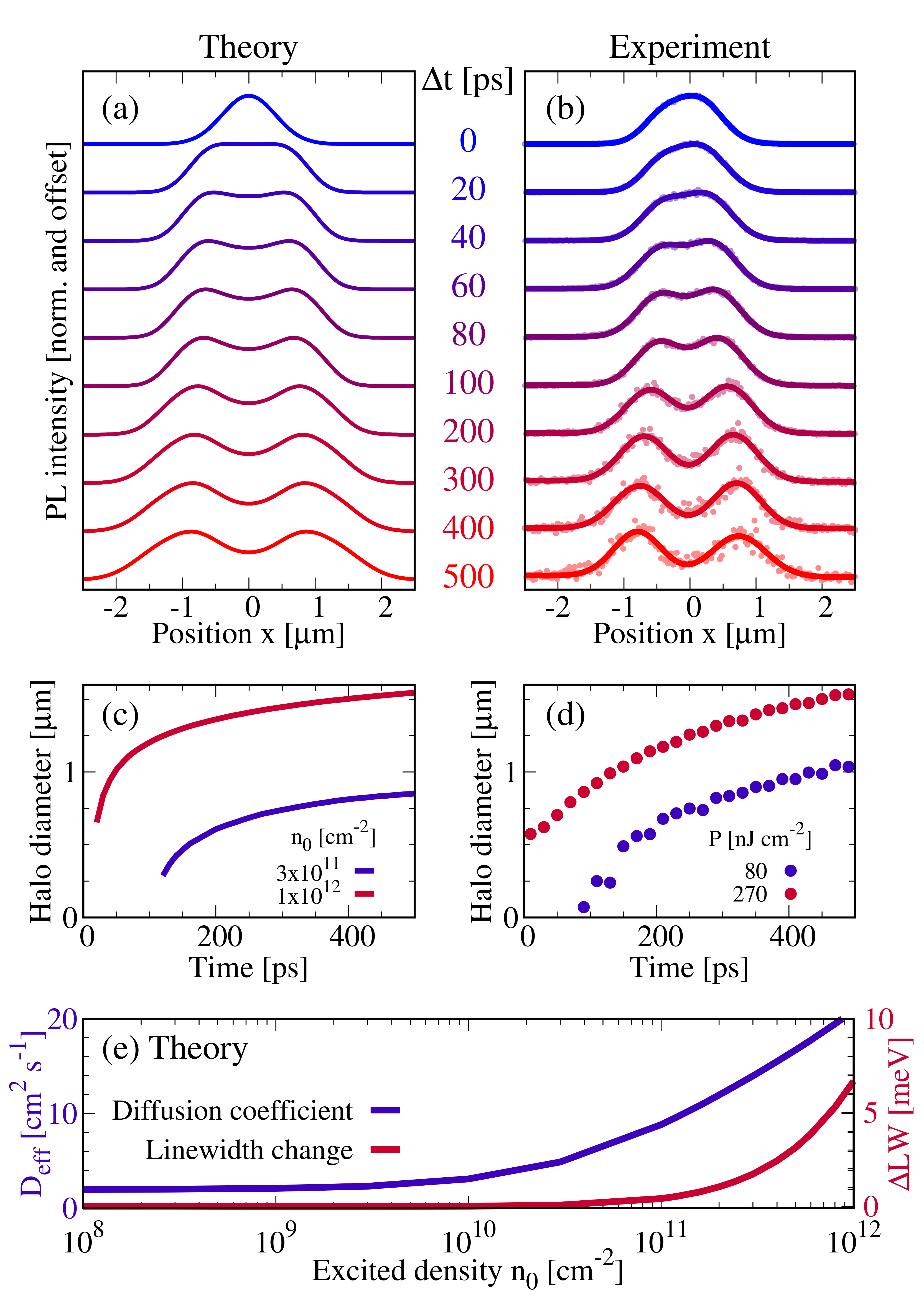}
 \vspace*{-0.7cm}
 \caption{\textbf{Direct theory-experiment comparison.} (a)-(b) Theoretically predicted and  experimentally measured temporal and spatial evolution of photoluminescence for an initially excited exciton density of $10^{12}\ \text{cm}^{-2}$  and an excitation energy density of 270\,nJ/cm$^2$ per pulse, respectively. The experimental data (circles) is presented together with double-Gaussian fit profiles (lines).
 (c)-(d) Theory-experiment comparison of the halo formation for two initial exciton densities. (e) Theoretical prediction of density dependence for the effective diffusion coefficient (blue line) and the linewidth change at the excitation spot evaluated 100 ps after the excitation (red line).
}
 \label{fig:experiment}
\end{figure}

\noindent\textbf{Theoretical and experimental photoluminescence}\\
Assuming a thermal equilibrium in the exciton distribution and determining the exciton occupation in the light cone\,\cite{selig2018dark}, we calculate the temporally and spatially resolved photoluminescence spectrum for WS$_2$ monolayers on a SiO$_2$ substrate at room temperature (Fig.\,\ref{fig:experiment} (a)). At the same time, we perform experimental PL measurements to be able to obtain a direct theory-experiment comparison (Fig.\,\ref{fig:experiment} (b)). More details on the experimental design can be found in the methods section. 
Figure~\ref{fig:experiment} shows an excellent qualitative agreement between theory and experiment. We observe a clear transition from a Gaussian to a super-Gaussian diffusion that leads to the formation of halos (Figs.~\ref{fig:experiment} (a)-(b)). 
We also obtain an equally good agreement, when it comes to the halo formation time that ranges on a timescale of tens to hundreds of picoseconds depending on the exciton density (Figs.~\ref{fig:experiment} (a)-(b)) as well as in the sub-linear time-dependence of the halo diameter (Figs.~\ref{fig:experiment} (c)-(d)). We find that the halos appear at earlier times for stronger excitations, since more excitons are present resulting in a strong hot-phonon effect and a larger thermal drift. Furthermore, we find that the halo diameter increases very quickly at the beginning, where excitons travel through a steep temperature gradient. Note that no parameters were fitted to obtain a quantitative agreement between theory and experiment. 

We experimentally observe a generally slower diffusion that is manifested in a narrower spatial distribution compared to the theoretical prediction (Figs.~\ref{fig:experiment} (a)-(b)). This can be ascribed to the scattering either with defects\,\cite{yuan2017exciton} or disorder, which has not been taken into account in the theoretical model. The reduced diffusion coefficient caused by such imperfections also implies that a smaller temperature gradient and thus a weaker thermal drift are sufficient to overcome the diffusion force and create halos. This also explains that halos are observed already at smaller exciton densities in the experiment. 

Finally, we predict exciton diffusion for different initial densities and extract the corresponding effective diffusion coefficients $D_{\text{eff}}$ from the slope of the variance of the spatial profiles. We find that $D_{\text{eff}}$ remains almost constant at low densities, where thermal effects and Auger scattering are not significant enough and is thus limited by the linear diffusion. Then, it strongly increases for higher excitation densities (Fig.~\ref{fig:experiment} (e)) -- in agreement with previous experimental observations\,\cite{kulig2018exciton}.
The effective increase of the diffusion coefficient occurs for three reasons: (1) the additional current corresponding to the thermal drift of excitons out of the hot region, (2) the larger diffusion coefficient arising from a higher excitonic temperature (larger average squared velocity) with a constant lattice temperature (i.e. scattering with acoustic phonons does not become stronger), and (3) Auger recombination that makes the distribution effectively broader.

The optical phonon density is also significantly affected by the excitation density (cf. Fig.~\ref{fig:nonlinearDiffusion} (c)) resulting from the relaxation of high-energy Auger-scattered excitons. The phonon density has a direct effect on the excitonic linewidth in optical spectra through changes in exciton-phonon scattering rates\,\cite{selig2016excitonic, brem2019intrinsic}. Figure~\ref{fig:experiment} (e) shows the predicted change in the excitonic linewidth at the excited spot due to hot optical phonons. We find that the linewidth change is negligibly small for initial exciton densities up to $10^{11}\ \text{cm}^{-2}$, however, it increases exponentially at higher densities. We  hope that our work will trigger future experimental studies investigating the spatial profile of the linewidth change.
\\

\noindent\textbf{\large{Discussion}}\\[5pt]
We study the diffusion of excitons with an initial density that is determined by an  excitation pulse, which is characterized by a Gaussian spatial, spectral and temporal profile.  
Since we focus  on the diffusion, we do not resolve the process of the optical excitation itself, but we assume an initial distribution of thermalized excitons, which can be described by a Boltzmann distribution. This can be justified by the fact that the optical excitation, exciton formation and thermalization occur on a much faster, sub-picosecond timescale\,\cite{Ceballos2016, Steinleitner2017} than slower diffusion processes\,\cite{jago2019spatio}. Furthermore, we focus on the diffusion of the most occupied exciton state, which in the case of tungsten disulfide (WS$_2$) is the momentum-dark K{\textLambda} exciton\,\cite{malic2018dark, deilmann2019finite, selig2016excitonic, selig2018dark, berghauser2018mapping, maja_sensor}. For the exciton population in a quasi-equilibrium between bright and dark states due to very fast inter-valley scattering rates compared to exciton lifetimes, the emission from the bright state should reflect the dynamics of the total population.
A more detailed study of the interplay of bright and dark excitonic states that might play additional role at lower temperatures and faster timescales  is beyond the scope of this work. Finally, we  focus on exciton diffusion and neglect phonon propagation effects, since the velocity of acoustic phonons is two orders of magnitude smaller than the average exciton velocity at room temperature\,\cite{jin2014intrinsic, glazov2019phonon}. Note that there is a recent theoretical study focusing on the formation of halos as a consequence of ballistic and diffusive phonon drag propagation effects\,\cite{glazov2019phonon}.

In summary, we have resolved the spatio-temporal dynamics of excitons in TMDs based on a microscopic equation of motion approach and supported by photoluminescence measurements. We find that at low excitations, exciton diffusion follows the standard Fick's law and can be described by a Gaussian spatial profile. We determined a diffusion coefficient of excitons in a defect-free WS$_2$ monolayer on a SiO$_2$ substrate to be in the range of $2\,\text{cm}^2\text{s}^{-1}$ at room temperature. More importantly, we revealed that at high excitations hot optical phonons give rise to a long-lived spatial gradient in the excitonic temperature. This results in pronounced thermal currents of excitons giving rise to super-Gaussian spatial profiles and the formation of spatial rings (halos) - in  excellent agreement with experimental observations. Finally, we predicted a significant increase of the diffusion coefficient and the excitonic linewidth with the excitation density due to  more efficient thermal currents. The obtained microscopic insights allow us to significantly advance in the fundamental understanding of exciton transport and its interplay with optics and exciton dynamics, which is of central importance for the realization of novel optoelectronics devices.\\

\noindent\textbf{\large{Materials and Methods}}\\[5pt]
\textit{Experimental design:}
 Spatially- and time-resolved emission microscopy\,\cite{kulig2018exciton} is performed on WS$_2$ monolayer flakes, exfoliated on polymer and stamped onto 290\,nm SiO$_2$/Si substrates\,\cite{Castellanos-Gomez2014a}. A pulsed Ti:sapphire laser with a repetition rate of 80\,MHz and 100\,fs pulse length is used for excitation. The laser is focused by a 100x microscope objective to a spot with a full width half maximum diameter of 0.5\,$\mu$m. The photon energy is tuned to 2.43\,eV, where the effective absorption coefficient of the sample on the particular substrate is about 0.1. A cross section in the middle of the photoluminescence spot is directly imaged onto a streak camera detector to simultaneously obtain spatially- and time-resolved profiles of the emission. \\
 \textit{Theoretical approach:}
To represent quantum-mechanical quantities simultaneously in real and momentum space, 
we introduce the Wigner function 
$
f_{\mathbf{k}}(\mathbf{r}) = \sum_{\mathbf{q}} e^{i \mathbf{q} \cdot \mathbf{r}} 
\big< a^{\dagger}_{\mathbf{k}-\frac{1}{2}\mathbf{q}} a^{\phantom{\dagger}}_{ \mathbf{k}+\frac{1}{2}\mathbf{q}} \big>
$
with $a^{\dagger}_{\mathbf{k}}$ and $a^{\phantom{\dagger}}_{\mathbf{k}}$ being creation and annihilation operators of particles with the momentum $\mathbf{k}$\,\cite{hess1996maxwell, rossi2002theory, jago2019spatio, wigner1932quantum}.
Exploiting the density matrix formalism, equations of motion for the spatio-temporal dynamics of excitons
$N_{\mathbf{Q}}(\mathbf{r},t) = \sum_{\mathbf{q}} e^{i \mathbf{q} \cdot \mathbf{r}} \big< X^{\dagger}_{\mathbf{Q} - \frac{1}{2}\mathbf{q}} X^{\phantom{\dagger}}_{\mathbf{Q} + \frac{1}{2}\mathbf{q}} \big>$
 and phonons 
 $
n_{\mathbf{q}}(\mathbf{r}, t) = \sum_{\mathbf{k}} e^{i \mathbf{k} \cdot \mathbf{r}} \big< b^{\dagger}_{\mathbf{q} - \frac{1}{2}\mathbf{k}} b^{\phantom{\dagger}}_{\mathbf{q} + \frac{1}{2}\mathbf{k}} \big>
$
  are derived. Here, $X_{\mathbf{Q}}^{\dagger}$ and $X_{\mathbf{Q}}$ are creation and annihilation operators for excitons with the momentum $\mathbf{Q}$,  while  $b_{\mathbf{q}}^{\dagger}$ and $b_{\mathbf{q}}$ are the respective operators for phonon creation and annihilation\,\cite{katsch2018theory, kira2011semiconductor}.


%

\noindent\textbf{Acknowledgments}\\[5pt]
This project has received funding from the European Union’s Horizon 2020 research and innovation programme under grant agreement No. 785219 (Graphene Flagship). Furthermore, we acknowledge support from the Swedish Research Council (VR). The Regensburg team further acknowledges the DFG for financial support via Emmy Noether Grant CH 1672/1-1 and Collaborative Research Center SFB 1277 (B05). Finally, we thank M. Glazov for helpful discussions.

\end{document}